\documentclass[12pt]{article}
\usepackage{a41}
\usepackage{epsfig}
\usepackage{axodraw}
\usepackage{amssymb}
\usepackage{amsmath}
\newcommand\e{{\rm e}}
\newcommand\PP{{\mathbb{P}}}

\newcommand{\bea}{\begin{eqnarray}}
\newcommand{\bq}{\begin{equation}}
\newcommand{\eea}{\end{eqnarray}}
\newcommand{\eq}{\end{equation}}

\newcommand{\Df}{\underset{\rm Def}{=}}

\newcommand\pvec{\mbox{\boldmath $p$}}
\newcommand\lvec{\mbox{\boldmath $l$}}

\newcommand\xx{\tilde{x}}

\begin{document}
\noindent
\sloppy
\thispagestyle{empty}
\begin{flushleft}
DESY 01--072 \hfill
{\tt hep-ph/0106037}\\
May   2001
\end{flushleft}
%
\vspace*{\fill}
\begin{center}
{\LARGE\bf  On the Scaling Violations of Diffractive}

\vspace{2mm}
{\LARGE\bf  Structure Functions: Operator Approach}

\vspace{2cm}
\large
Johannes Bl\"umlein$^a$  and
Dieter Robaschik$^{a,b}$
\\
\vspace{2em}
\normalsize
{\it $^a$~Deutsches Elektronen--Synchrotron, DESY,\\
Platanenallee 6, D--15738 Zeuthen, Germany}
\\
\vspace{2em}
{\it $^b$~Brandenburgische Technische Universit\"at Cottbus, 
Fakult\"at 1,}\\
{\it  PF 101344, D--03013  Cottbus, Germany} \\
\end{center}
\vspace*{\fill}
%
\begin{abstract}
\noindent
A quantum field theoretic treatment of inclusive deep--inelastic 
diffractive scattering is given. The process can be described in the
general framework of non--forward scattering processes using the 
light--cone expansion in the generalized Bjorken region. Evolution 
equations of the diffractive hadronic matrix elements are derived at the 
level of the twist--2 contributions and are compared to those of inclusive 
deep--inelastic forward scattering (DIS). The diffractive parton densities 
are obtained as projections of two--variable parton distributions. We 
also comment on the higher twist contributions in the light--cone 
expansion.
\end{abstract}
\vspace*{\fill}
\newpage
\section{Introduction}
\label{sec-1}

\vspace{1mm}
\noindent
Deep inelastic diffractive lepton--nucleon scattering was observed at the
electron--proton collider HERA some years ago~\cite{EXP1}. This process 
is measured in detail by now~\cite{EXP2} and the structure function 
$F_2^D(x,Q^2)$ was extracted\footnote{The measurement of the longitudinal
diffractive structure function $F_L^D(x,Q^2)$ has not yet been possible. 
For the DIS structure function cf.~\cite{FL}.} The diffractive events are
characterized by a rapidity gap between the final state nucleon and the 
set of the diffractively produced hadrons due to a color--neutral 
exchange. The (semi-inclusive) structure functions emerging in the 
diffractive scattering cross section show the same scaling violations as 
the structure functions in deep inelastic scattering within the current 
experimental resolution~\cite{EXP3}. This is a remarkable fact, which 
should also be understood with the help of field--theoretic methods within 
Quantum Chromodynamics (QCD). In the description of diffractive $ep$ 
scattering one has to clearly distinguish~\cite{BAKO} the case of hard 
scattering, i.e. large $Q^2$, from that of softer hadronic interactions, 
cf. also~\cite{DL}. A perturbative description of the scaling violations 
can only by hoped for in the former case, to which we limit our 
considerations in this paper.

There is a vast amount of varying descriptions of the underlying dynamics
of the diffractive scattering process\footnote{For surveys and a recent
account see~\cite{SUR,NP}.}, ranging from intuitive phenomenological 
models~\cite{INGEL} to non-perturbative semi-classical 
descriptions~\cite{BH}, being able to describe and to parameterize 
\cite{EBKW} the existing data differentially.

In this paper we describe the process of inclusive deep--inelastic
diffractive scattering, being a non--forward process, at large space--like 
momentum transfer using the general light--cone expansion for the 
non--forward case, cf.~[12--14]. As will be shown, the scaling
violations of 
the diffractive structure functions can be described perturbatively in 
this region. We show that the diffractive parton densities can be derived
as a special projection of  two--variable distribution functions 
$f^A(z_+,z_-)$, where $z_+$ and $z_-$ are light--cone momentum fractions;
for other projections in similar cases related to different observables 
see e.g.~[13--15]. In particular we have {\it no need} to
refer to the specific mechanism of the non--perturbative color--singlet 
exchange between the proton and the set of diffractively produced hadrons
and consider instead the expectation values of operators of a given twist
and their anomalous dimensions. In this way we can generalize the analysis 
to operators of higher twist. This formulation is very appropriate for 
potential later studies of the corresponding operator matrix elements with 
the help of lattice techniques similar to the case of deep--inelastic 
scattering \cite{LAT} calculating their Mellin--moments.

The paper is organized as follows. In section~2 we consider the Lorentz
structure of the differential cross section for inclusive deep--inelastic 
diffractive scattering. The short--distance structure of the matrix 
element is derived in section~3 using the (non--local) light--cone 
expansion in the generalized Bjorken region. In section~4 we derive the 
anomalous  dimensions for the case of twist--2, discuss the implications 
for higher twist operators and compare the case of diffractive scattering
to that of deep--inelastic scattering, and section~5 contains the
conclusions.

\section{Lorentz Structure}
\vspace*{-5mm}\noindent
\label{sec-2}

\vspace{1mm}
\noindent
The process of deep--inelastic diffractive scattering is described by the
diagram Figure~1.
The differential scattering cross section for single--photon 
exchange is given by
\begin{equation}
\label{eqD1}
d^5 \sigma_{\rm diffr}
= \frac{1}{2(s-M^2)} \frac{1}{4} dPS^{(3)} \sum_{\rm spins}
\frac{e^4}{Q^2} L_{\mu\nu} W^{\mu\nu}~.
\end{equation}
Here $s=(p_1+l)^2$ is the cms energy of the process squared and $M$ 
denotes the nucleon mass.

\vspace*{0.8cm}
\begin{center}
\begin{picture}(-50,100)(0,0)
\setlength{\unitlength}{0.2mm}
\SetWidth{1.5}
\ArrowLine(-150,100)(-100,100)
\ArrowLine(-100,100)(-50,120)
\Photon(-100,100)(-70,70){5}{5}%
\ArrowLine(-30,30)(0,0)
\ArrowLine(-100,0)(-70,30)
\SetWidth{5}
\ArrowLine(-30,70)(0,100)
\SetWidth{1.5}
\CCirc(-50,50){26}{Black}{Yellow}
\setlength{\unitlength}{1pt}
\Text(-160,110)[]{$l$}
\Text(-40,130)[]{$l'$}
\Text(-90,75)[]{$q$}
\Text(-110,-10)[]{$p_1$}
\Text(10,-10)[]{$p_2$}
\Text(10,110)[]{$M_X$}
\end{picture}
\end{center}

\vspace*{1mm}
\noindent
\begin{center}
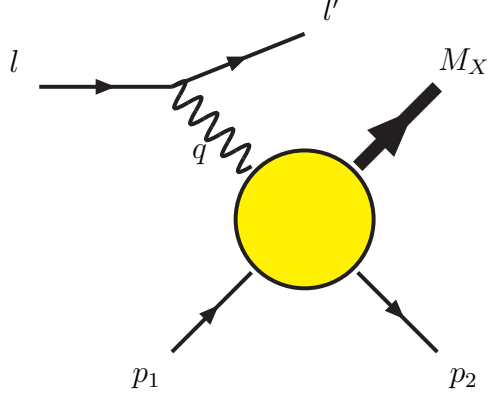
{\sf Figure~1:~The virtual photon-hadron amplitude for 
diffractive $ep$ scattering} 
\end{center}
The phase space $dPS^{(3)}$ depends on five variables\footnote{For the
commonly used notation of four variables see e.g.~\cite{MD}.} since one 
final state mass varies. They can be chosen as 
Bjorken~$x= Q^2/(W^2+Q^2-M^2)$, the photon virtuality $Q^2=-q^2$, 
$t = (p_1 - p_2)^2$, a variable describing the non--forwardness w.r.t.
the incoming proton direction,
\begin{equation}
\label{eqV1}
x_{\PP} = - \frac{2\eta}{1-\eta} = 
\frac{Q^2 + M_X^2 - t}{Q^2 + W^2 -M^2} \geq  x~,
\end{equation}
demanding $M_X^2 > t$ and where
\begin{equation}
\label{eqV2}
\eta = \frac{q.(p_2 - p_1)}{q.(p_2+p_1)}~\epsilon~\left[-1,\frac{-x}{2-x}
\right]~,
\end{equation}
and $\Phi$ the angle between the lepton plane $\pvec_1 \times \lvec$ and
the hadron plane $\pvec_1 \times \pvec_2$,
\begin{equation}
\label{eqV3}
\cos \Phi = \frac{(\pvec_1 \times \lvec).(\pvec_1 \times  \pvec_2)}
                 {|\pvec_1 \times \lvec ||\pvec_1 \times  \pvec_2|}~.
\end{equation}
$W^2 = (p_1+q)^2$ and  $M_X^2 = (p_1+q-p_2)^2$ denote the hadronic
mass squared and the square of the diffractive mass, respectively.

Since the leptonic tensor $L^{\mu\nu}$ is symmetric for unpolarized
scattering and the 
electromagnetic current is conserved the  hadronic tensor, which is
generally composed out of the tensors $g_{\mu\nu}, (q_\mu, p_{1\mu},
p_{2\mu}) \otimes (q_\nu, p_{1\nu}, p_{2\nu})$, consist out of four
structure functions only\footnote{For a construction method, see e.g.
\cite{TARR}.}
\begin{eqnarray}
\label{eqD2}
W_{\mu\nu} &=& \left( -g_{\mu\nu} + \frac{q_{\mu}q_{\nu}}{q^2}\right) W_1
           + \left( p_{1\mu} - q_\mu \frac{p_1.q}{q^2}\right)
             \left( p_{1\nu} - q_\nu \frac{p_1.q}{q^2}\right) 
             \frac{W_3}{M^2}
\nonumber\\
           & &~~~~~~~~~~~~~~~~~~~~~~~~~+ \left(
p_{2\mu} - q_\mu \frac{p_2.q}{q^2}\right)
             \left( p_{2\nu} - q_\nu \frac{p_2.q}{q^2}\right) 
             \frac{W_4}{M^2}
\nonumber\\
          &+& \left[\left(p_{1\mu} - q_\mu \frac{p_1.q}{q^2}\right)
                   \left(p_{2\nu} - q_\nu \frac{p_2.q}{q^2}\right)
           +       \left(p_{2\mu} - q_\mu \frac{p_2.q}{q^2}\right)
                   \left(p_{1\nu} - q_\nu \frac{p_1.q}{q^2}\right)
\right]  \frac{W_5}{M^2}~,
\end{eqnarray}
with
\begin{equation}
\label{eqD3}
W_i = W_i(x,Q^2,x_{\PP},t)~.
\end{equation}
The size of the rapidity--gap \cite{COL,NP} is of the order
$\Delta y \sim \ln(1/x_{\PP})$ and large in the kinematic domain of HERA.

If target masses can be neglected, $M^2 \sim 0$, and $t$ is considered
to be very small compared to all other invariants the in-- and outgoing
proton four--momenta become proportional,~$p_2 = z p_1$. The variable
$z$ is then related to $x_{\PP}$ and $\eta$ by
\begin{equation}
\label{eqD4}
z = 1 - x_{\PP} =  \frac{1 + \eta}{1 - \eta}~.
\end{equation}
In this limit the hadronic tensor is described by two structure functions
\begin{eqnarray}
\label{eqD5}
W_{\mu\nu} &=& \left( -g_{\mu\nu} + \frac{q_{\mu}q_{\nu}}{q^2}\right) W_1
           + \left( p_{1\mu} - q_\mu \frac{p_1.q}{q^2}\right)
             \left( p_{1\nu} - q_\nu \frac{p_1.q}{q^2}\right) 
             \frac{W_2}{M^2}~,
\end{eqnarray}
with
\begin{eqnarray}
\label{eqD6}
W_2 = W_3 + (1-x_{\PP}) W_5 + (1-x_{\PP})^2 W_4~.
\end{eqnarray}
Due to the dependence on $x_{\PP}$ or $\eta$,~Eq.~(\ref{eqV1}), the
process is {\it non--forward} although the algebraic structure of the
hadronic tensor is the same as in the forward case. Finally the 
generalized Bjorken--limit is carried out, 
\begin{eqnarray}
\label{eqBL}
2 p_1.q = 2M \nu \rightarrow \infty,~~~~~p_2.q \rightarrow \infty,~~~~~
Q^2 \rightarrow \infty~~~{\rm with}~~x~~{\rm and}~~x_{\PP} =~{\rm fixed}~,
\end{eqnarray}
which leads to
$M W_1(x,Q^2,x_{\PP}) \rightarrow F_1(x,Q^2,x_{\PP})$ and
$\nu W_2(x,Q^2,x_{\PP}) \rightarrow F_2(x,Q^2,x_{\PP})$.
Note that $Q^2 \gg t, M^2$ always holds in this region.
In the limit $p_2 = zp_1$ and $M^2,t = 0$ the $\Phi-$integral becomes 
trivial. The scattering cross section reads
\begin{eqnarray}
\label{eqD7}
\frac{d^3 \sigma_{\rm diffr}}{dx dQ^2 d x_{\PP}} = \frac{2 \pi \alpha^2}
{x Q^4} \left[y^2 2x F_1^{D(3)}(x,Q^2,x_{\PP})
+ 2(1-y) F_2^{D(3)}(x,Q^2,x_{\PP})\right]~,
\end{eqnarray}
with $y = q.p_1/l.p_1$.
\section{The Compton Amplitude}
\label{sec-3}

\vspace{1mm}
\noindent
The renormalized and time--ordered product of two electromagnetic
currents is given by
\begin{eqnarray}
\widehat{T}_{\mu\nu}(x) &=&
i RT \left[J_\mu\left(\frac{x}{2}\right)J_\nu\left(-\frac{x}{2}\right) S 
\right] \nonumber\\
&=&
 -e^2 \frac{\tilde x^\lambda}{2 \pi^2 (x^2-i\epsilon)^2}
 RT
 \left[
\overline{\psi}
\left(\frac{\tilde x}{2}\right)
\gamma^\mu \gamma^\lambda \gamma^\nu \psi
\left(-\frac{\tilde x}{2}\right)
- \overline{\psi}
\left(-\frac{\tilde x}{2}\right)
\gamma^\mu \gamma^\lambda \gamma^\nu \psi
\left(\frac{\tilde x}{2}\right)
\right] S
\end{eqnarray}
$\tilde x$ denotes a light--like vector corresponding to $x$,
\begin{eqnarray}
\label{xtil}
\tilde x = x + \frac{\zeta}{\zeta^2}\left[ \sqrt{x.\zeta^2 - x^2 \zeta^2}
- x.\zeta\right]~,
\end{eqnarray}
and $\zeta$ is a subsidiary vector. Following Refs.~\cite{BGR,BR} the
operator $\widehat{T}_{\mu\nu}$ can be expressed in terms of a vector
and an axial--vector operator by
\begin{eqnarray}
 \widehat{T}_{\mu\nu}(x)  =
 - e^2 \frac{\tilde x^\lambda}{i \pi^2 (x^2-i\epsilon)^2}
 \left[S_{\alpha \mu\lambda \nu} 
 O^\alpha\left(\frac{\tilde x}{2}, -\frac{\tilde x}{2}\right)
+i \varepsilon_{\mu\lambda \nu \sigma} 
O_5^\alpha  \left(\frac{\tilde x}{2}, -\frac{\tilde x}{2}\right)
\right]~,
\end{eqnarray}
where
\begin{eqnarray}
S_{\alpha \mu\lambda \nu} = g_{\alpha\mu}g_{\lambda \nu}
                          + g_{\lambda\mu}g_{\alpha \nu}
                          - g_{\mu\nu}g_{\lambda \alpha}~.
\end{eqnarray}
The  bilocal light--ray operators are
\begin{eqnarray}
\label{oo}
O^{\alpha}\left(\frac{\tilde x}{2},-\frac{\tilde x}{2}\right)
&=&
\frac{i}{2}RT
\left[\overline{\psi}\left(\frac{\tilde x}{2}\right)
\gamma^\alpha\psi\left(-\frac{\tilde x}{2}\right)
- \overline{\psi}\left(-\frac{\tilde x}{2}\right)
\gamma^\alpha\psi\left(\frac{\tilde x}{2}\right)\right]S~,
\\
\label{oo5}
O^{\alpha}_5\left(\frac{\tilde x}{2},-\frac{\tilde x}{2}\right)
&=&
\frac{i}{2} RT
\left[\overline{\psi}\left(\frac{\tilde x}{2}\right)
\gamma_5\gamma^\alpha\psi\left(-\frac{\tilde x}{2}\right)
+ \overline{\psi}\left(-\frac{\tilde x}{2}\right)
\gamma_5\gamma^\alpha\psi\left(\frac{\tilde x}{2}\right)\right]S~.
\end{eqnarray}
The general operator $\widehat{T}_{\mu\nu}$ is now to be related to the
diffractive scattering cross section. This is possible upon applying
Mueller's generalized optical theorem~\cite{AHM} (Figure~2), which
moves the final state proton into an initial state anti-proton.

\vspace*{7mm}
\begin{center}
\begin{picture}(200,100)(0,0)
\setlength{\unitlength}{0.2mm}
\SetWidth{1.5}
\Line(-107,102)(-107,-2)
\Line(7,102)(7,-2)
\Photon(-100,100)(-70,70){5}{5}
\ArrowLine(-30,30)(0,0)
\ArrowLine(-100,0)(-70,30)
\SetWidth{5}
\ArrowLine(-30,70)(0,100)
\SetWidth{1.5}
\CCirc(-50,50){26}{Black}{Yellow}
\setlength{\unitlength}{1pt}
\Text(15,107)[]{\large $2$}
\Text(50,50)[]{\large $=~~{\rm Disc}$}
\Text(-105,115)[]{$q$}
\Text(-105,-15)[]{$p_1$}
\Text(5,-15)[]{$p_2$}
\Text(75,115)[]{$q$}
\Text(75,-15)[]{$p_1$}
\Text(95,-15)[]{$p_2$}
\Text(60,35)[]{$X$}
\Text(175,20)[]{$X$}
\Text(275,115)[]{$q$}
\Text(275,-15)[]{$p_1$}
\Text(255,-15)[]{$p_2$}
\setlength{\unitlength}{0.2mm}
\SetWidth{1.5}
\Photon(80,100)(110,70){5}{5}
\ArrowLine(80,0)(110,30)
\ArrowLine(130,30)(100,0)
\Line(150,50)(200,50)
\Line(150,45)(200,45)
\Line(150,55)(200,55)
\Line(150,40)(200,40)
\Line(150,60)(200,60)
\ArrowLine(240,30)(270,0)
\ArrowLine(250,0)(220,30)
\CCirc(130,50){26}{Black}{Yellow}
\CCirc(220,50){26}{Black}{Yellow}
\Photon(240,70)(270,100){5}{5}
\end{picture}
\end{center}

\vspace*{10mm}\noindent
\begin{center}
{\sf Figure~2:~A. Mueller's optical theorem.}
\end{center}
Consequently, the Compton amplitude is obtained as the expectation value
\begin{eqnarray}
T_{\mu\nu}(x) = \langle p_1,-p_2|\widehat{T}_{\mu\nu}|p_1,-p_2\rangle~,
\end{eqnarray}
which is forward w.r.t. to the direction defined by $\langle p_1,-p_2|$.
The quantity $\langle p_1,-p_2|$, in the strict sense, is not a {\it
single} physical state of the Hilbert-space of states, since this would 
imply
$t=(p_2 -p_1)^2 > 0$. The analytic continuation is such, that $t$ is
kept space--like, which is necessary for the non--perturbative
behavior of the matrix element. 
The twist--2 contributions to the expectation values of the operators 
(\ref{oo},\ref{oo5})  are obtained
\begin{eqnarray}
\label{eqSCA}
\langle p_1,-p_2|O^{A,\mu}_{(5)}(\kappa_+ \xx,\kappa_-\xx)|
p_1,-p_2\rangle = \left.
\int_0^1 d\lambda
\partial^{\mu}_x
\langle p_1,-p_2|O_{(5)}^A(\lambda \kappa_+ x,\lambda \kappa_-x)
|p_1,-p_2\rangle \right|_{x=\xx}
\end{eqnarray}
as partial derivatives of the expectation values of
\begin{eqnarray}
O^A(\kappa_+x,\kappa_-x)
= x^\alpha O^A_\alpha(\kappa_+x,\kappa_-x)
\end{eqnarray}
the corresponding scalar and pseudo-scalar operators. The index $A = q,G$
labels the quark-- or gluon operators, cf.~\cite{BGR}. In the unpolarized 
case only the vector operators contribute. The scalar twist--2 quark
operator
matrix element has the representation\footnote{For parameterizing hadronic
matrix elements see e.g.~\cite{PARA}.} 
due to the overall symmetry in $x$
\begin{eqnarray}
\label{eqSCAM}
\langle p_1,-p_2|O^q(\kappa_+ x,\kappa_- x)|p_1,-p_2\rangle = xp_-
\!\!\!         \int \!\!\!
Dz ~\e^{-i \kappa_- x p_z} {f}^q(z_+,z_-)
+ x\pi_-\!\!\!         \int \!\!\!
Dz ~\e^{-i \kappa_- x p_z} {f}^q_\pi(z_+,z_-),
\end{eqnarray}
and is independent of $\kappa_+$. Here
we assume that all the trace--terms have been subtracted, see
\cite{BGR}.
The action of $\langle p_1,-p_2|$ onto the scalar operator yield
projections onto two directions $p_-$ and $\pi_- = p_+ - p_-/\eta$.
${f}^A(z_+,z_-)$ and $f^A_\pi(z_+,z_-)$ denote the corresponding
scalar 
two--variable distribution amplitudes and the measure $Dz$ is
\begin{eqnarray}
\label{Dz}
Dz = d z_+ d z_-
\theta(1 + z_+ + z_-) \theta(1 + z_+ - z_-)
\theta(1 - z_+ + z_-) \theta(1 - z_+ - z_-)~.
\end{eqnarray}
Here, we decomposed the vector $p_z$ as
\begin{eqnarray}
p_z = p_- z_- + p_+ z_+ =  p_- \vartheta + \pi_- z_+~,
\end{eqnarray}
with $z_{1,2}$ momentum fractions along $p_{1,2}$ and $p_{\pm} = p_2 \pm
p_1$, $z_\pm = (z_2 \pm z_1)/2$ and
\begin{eqnarray}
\vartheta = z_- + \frac{1}{\eta} z_+.
\end{eqnarray}
Note that, $q.\pi_- \equiv 0$. In the approximation $M^2, t \sim 0$, in 
which we work from now on, the vector $\pi_-$ vanishes.
In this limit only the first term contributes to the matrix element
Eq.~(\ref{eqSCAM}).

The Fourier--transform of the Compton amplitude is given by~\cite{BR}
\begin{eqnarray}
T_{\mu\nu}(p_1,p_2,q) &=& i \int d^4x \e^{iqx} T_{\mu\nu}(x) \nonumber\\
&=& - 2 S_{\rho\mu\sigma\nu} \int Dz \left[
\frac{p_-^{\rho} Q_z^{\sigma}}{Q^2_z + i\varepsilon} - \frac{1}{2}
\frac{p_z^{\rho} p_-^{\sigma}}{Q^2_z + i\varepsilon}%
+ \frac{Q_z.p_-}{(Q^2_z+i\varepsilon)^2}
p_z^{\rho} Q_z^{\sigma}\right] F(z_+,z_-)~,
\end{eqnarray}
with $Q_z = q - p_z/2$. The function $F(z_+,z_-)$ is related to the
distribution function $f(z_+,z_-)$ by
\begin{eqnarray}
F(z_+,z_-) = \int_0^1 \frac{d\lambda}{\lambda^2} 
f\left(\frac{z_+}{\lambda},\frac{z_-}{\lambda}\right)
\theta(\lambda-|z_+|)\theta(\lambda-|z_-|)~.
\end{eqnarray}
The denominators take the form
\begin{eqnarray}
\frac{1}{Q^2_z+i\varepsilon} = - \frac{1}{qp_-}\frac{1}{(\vartheta
- 2\beta  + i \varepsilon)}~,
\end{eqnarray}
where
\begin{eqnarray}
\beta =  \frac{x}{x_{\PP}} = \frac{q^2}{2q.p_-}~.
\end{eqnarray}
The conservation of the electromagnetic current is easily seen
\begin{eqnarray}
q^{\mu}T_{\mu\nu}(p_1,p_2,q)  =
q^\nu T_{\mu\nu}(p_1,p_2,q)  =
- 2 p_{-,\nu(\mu)} \int Dz F(z_+,z_-) = 0~,
\end{eqnarray}
since the symmetry relation, \cite{BR},
\begin{eqnarray}
\int Dz F(z_+,z_-) = 0~,  
\end{eqnarray}
holds.
Subsequently we will use the distribution
\begin{eqnarray}
\label{eqFF}
\widehat{F}(\vartheta,\eta) = \int Dz F(z_+,z_-) 
\delta(\vartheta - z_- -z_+/\eta) = \int_\vartheta^{-{\rm sign}(\vartheta)
/\eta} \frac{dz}{z} \widehat{f}(z,\eta)~.
\end{eqnarray}
The distribution function $\widehat{f}(z,\eta)$ is related to 
$f(z_+,z_-)$ by
\begin{eqnarray}
\label{eqFF1}
\widehat{f}(z,\eta) = \int^{\eta(1-z)}_{\eta(1+z)} d\rho~
\theta(1-\rho) \theta(\rho+1) f(\rho,z-\rho/\eta)~,
\end{eqnarray}
with $\rho = z_+/\eta$.

The Compton amplitude may be re-written as
\begin{eqnarray}
\label{eqCOMP1}
T_{\mu\nu}(p_1,p_2,q) &=& 
-2 \int_{+1/\eta}^{-1/\eta} d\vartheta \Biggl\{
\left[-q.p_- g_{\mu\nu}
-2\vartheta p_{-,\mu} p_{-,\nu} + (p_{-,\mu}q_\nu +p_{-,\nu} q_\mu)\right]
\frac{1}{Q_z^2+i\varepsilon} \nonumber\\
& & ~~~~+ \left[\vartheta (q_\mu p_{-,\nu} + q_\nu p_{-,\mu}) - \vartheta
q.p_- g_{\mu\nu} - \vartheta^2 p_{-,\mu} p_{-,\nu} \right]\frac{1}
{(Q_z^2+i\varepsilon)^2}\Biggr\} \nonumber\\
& &~~~~~~~~~~~~~~~~~~~~~~~~~~~~~~~~~~~~~~~~~~~~~~~
\times \int_\vartheta^{-{\rm sign}(\vartheta)/\eta}\frac{dz}{z}
\widehat{f}(z,\eta)~.
\end{eqnarray}
The $\vartheta$--integrals in Eq.~(\ref{eqCOMP1}) can be simplified
using the identities
\begin{eqnarray}
\int_{+1/\eta}^{-1/\eta} d\vartheta \frac{\vartheta^k}
{(\vartheta-2\beta+i\varepsilon)^2}\int_\vartheta^{{\rm sign}(\vartheta)
/\eta} \frac{dz}{z} \widehat{f}(z,\eta)
&=&
\int_{+1/\eta}^{-1/\eta} d\vartheta \frac{k \vartheta^{k-1}}
{(\vartheta-2\beta+i\varepsilon)}\int_\vartheta^{{\rm sign}(\vartheta)
/\eta} \frac{dz}{z} \widehat{f}(z,\eta) \nonumber\\    & &
~~~~~~~-\int_{+1/\eta}^{-1/\eta} d\vartheta \frac{\vartheta^{k-1}
\widehat{f}(\vartheta,\eta)}
{(\vartheta-2\beta+i\varepsilon)}~.
\end{eqnarray}
One may further re-write Eq.~(\ref{eqCOMP1}) referring to $p_1$ instead
of $p_-$,
\begin{eqnarray}
\label{eqCOMP2}
T_{\mu\nu}(p_1,p_2,q) &=& 
2 \int_{+1/\eta}^{-1/\eta} d\vartheta
\left[-g_{\mu\nu} 
+ \frac{p_{1,\mu}q_\nu +p_{1\nu} q_\mu)}{q.p_1}
+\vartheta  x_{\PP} \frac{p_{1\mu} p_{1\nu}}{q.p_1}\right]
\frac{\widehat{f}(\vartheta,\eta)}{(\vartheta - 2\beta +i\varepsilon)}~.
\end{eqnarray}
Before we consider the absorptive part of the Compton amplitude
we exploit the symmetry relation for the unpolarized distribution 
functions  $F^A(z_1,z_2)$, Ref.~\cite{BR},
\begin{eqnarray}
\label{eqSYM1}
F^A(z_1,z_2) = - F^A(-z_1,-z_2)~.
\end{eqnarray}
It translates into
\begin{eqnarray}
\label{eqSYM2}
\widehat{F}^A(\vartheta,\eta) = - \widehat{F}^A(-\vartheta,\eta)~,
\end{eqnarray}
and, cf. Eq.~(\ref{eqFF}),
\begin{eqnarray}
\label{eqSYM3}
\widehat{f}^A(\vartheta,\eta) = - \widehat{f}^A(-\vartheta,\eta)~.
\end{eqnarray}
Seeking the form of Eq.~(\ref{eqD5}), Eq.~(\ref{eqCOMP2}) transforms to
\begin{eqnarray}
\label{eqCOMP3}
T_{\mu\nu}(p_1,p_2,q) &=& \int_{1/\eta}^{-1/\eta} d\vartheta \left[
\left(-g_{\mu\nu} +\frac{q_\mu q_\nu}{q^2}\right) + \frac{2x}{q.p_1}
\left(p_{1\mu} - q_\mu \frac{p_1.q}{q^2}\right)
\left(p_{1\nu} - q_\nu \frac{p_1.q}{q^2}\right) \right] \nonumber\\
& &~~~~~~~~~~\times
\left[\frac{\widehat{f}(\vartheta,\eta)}
{\vartheta - 2 \beta + i\varepsilon}
-
\frac{\widehat{f}(-\vartheta,\eta)}
{\vartheta - 2 \beta + i\varepsilon}\right]\nonumber\\
& & ~~~~~~~~~~~+ 2\frac{p_{1\mu} p_{1\nu}}{q.p_1}
\int_{1/\eta}^{-1/\eta}
d\vartheta \left(\vartheta x_{\PP} - 2x
\right) \frac{\widehat{f}(\vartheta,\eta)}{\vartheta - 2\beta + 
i \varepsilon}~.
\end{eqnarray}
Taking the absorptive part one obtains
\begin{eqnarray}
\label{eqIMPAR}
W_{\mu\nu} &=& \frac{1}{2\pi}~{\sf Im}~T_{\mu\nu}(p_1,p_2,q) \\
           &=&       
\left(-g_{\mu\nu} +\frac{q_\mu q_\nu}{q^2}\right) 
{\sf F}_1(\beta,\eta,Q^2)
+ \frac{1}{q.p_1}
\left(p_{1\mu} - q_\mu \frac{p_1.q}{q^2}\right)
\left(p_{1\nu} - q_\nu \frac{p_1.q}{q^2}\right) 
{\sf F}_2(\beta,\eta,Q^2) \nonumber
\end{eqnarray}
where
\begin{eqnarray}
\label{eqF1}
{\sf F}_1(\beta,\eta,Q^2) = \sum_{q=1}^{N_f} 
e_q^2\left[f_q^D(\beta,Q^2,x_{\PP})+\overline{f}_q^D(\beta,Q^2,x_{\PP})
\right]
\equiv  F_1^{D(3)}(x,Q^2,x_{\PP})~,
\end{eqnarray}
with $N_f$ the number of flavors, choosing the factorization scale
$\mu^2 = Q^2$, and
\begin{eqnarray}
\label{eqCG}
{\sf F}_2(\beta,\eta,Q^2) = 2x {\sf F}_1(\beta,\eta,Q^2) \equiv
F_2^{D(3)}(x,Q^2,x_{\PP}).
\end{eqnarray}
If we compare the distribution functions $f_A^{D}$ with the parton
distributions in the deep inelastic case, the role of $x$ in the latter
case is taken by the variable $\beta$. Both variables have the range
$[0,1]$, through which Mellin transforms w.r.t. these variables can
be evaluated. On the contrary, the support in $x$ of the diffractive
structure functions is limited to $[0,x_{\PP}]$. Although $\beta$ in the
diffractive case compares to $x$ in the deep inelastic case, the 
Callan--Gross relation, at lowest order in $\alpha_s$, Eq.~(\ref{eqCG}), 
is given by the factor $2x$ between the two structure functions. 

The diffractive quark and anti--quark densities are given by
\begin{eqnarray}
\label{eqPART}
\sum_{q=1}^{N_f} e_q^2 f_q^D(\beta,Q^2,x_{\PP}) &=&
-\widehat{f}(2\beta,\eta,Q^2) \nonumber\\
\sum_{q=1}^{N_f} e_q^2 \overline{f}_q^D(\beta,Q^2,x_{\PP}) &=&
~~\widehat{f}(-2\beta,\eta,Q^2)~.
\end{eqnarray}
We finally can express the diffractive parton densities in terms of
the distribution function $f(z_+,z_-)$ directly
\begin{eqnarray}
\label{eqREL}
f^D(\beta,Q^2,x_{\PP}) &=& - \int_{-\frac{x_{\PP}+2x}{2-x_{\PP}}}
                              ^{-\frac{x_{\PP}-2x}{2-x_{\PP}}}
d \rho f(\rho,2\beta + \rho(2-x_{\PP})/x_{\PP},Q^2)\nonumber\\
\overline{f}^D(\beta,Q^2,x_{\PP}) &=& 
- \int_{-\frac{x_{\PP}+2x}{2-x_{\PP}}}
                              ^{-\frac{x_{\PP}-2x}{2-x_{\PP}}}
d \rho f(\rho,-2\beta + \rho(2-x_{\PP})/x_{\PP},Q^2)~.
\end{eqnarray}
\section{Evolution Equations}
\label{sec-4}

\vspace{1mm}
\noindent
The evolution equation of the scalar twist--2 quark and gluon operator 
Eq.~(\ref{eqSCA}) read, cf.~\cite{BGR},\footnote{The non--singlet
evolution equations are structurally the same as that with the
anomalous dimension $\gamma^{qq}$.}
\begin{eqnarray}
\mu^2 \frac{d}{d\mu^2} O^A(\kappa_+\xx, \kappa_-\xx; \mu^2) =
\int D\kappa' \gamma^{AB}(\kappa_+,\kappa_-,\kappa_+',\kappa_-';\mu^2)
O_B(\kappa_+'\xx,\kappa_-'\xx;\mu^2)~,
\end{eqnarray}
with $\kappa_\pm =(\kappa_2 - \kappa_1)/2$ and the measure
\begin{eqnarray}
D\kappa =   d\kappa_+ d\kappa_- \theta(1+\kappa_+ + \kappa_-)
                                \theta(1+\kappa_+ - \kappa_-)
                                \theta(1-\kappa_+ + \kappa_-)
                                \theta(1-\kappa_+ - \kappa_-)~.
\end{eqnarray}
We consider the non--forward evolution equations to trace an eventual
$\eta$--dependence.
Here $\gamma^{AB}(\kappa_+,\kappa_-,\kappa_+',\kappa_-',\mu^2)$ is the
general non--forward twist--2 singlet evolution 
matrix.~\footnote{The leading order and next--to--leading
order non--singlet and singlet anomalous dimensions were calculated in
Refs.~\cite{NFLO,RAD} and \cite{NFNLO}.}
The same evolution
equation applies to the matrix elements
$\langle p_1,-p_2|O^A(\kappa_+,\kappa_-)|p_1,-p_2\rangle$, which may be
related to the distribution functions $f^A(\vartheta,\eta)$ 
by\footnote{Note that     we identified here $\tau = \xx.p_-/\xx.p_+$
with $\eta = q.p_-/q.p_+$ which is possible after the Bjorken limit is
taken.}
\begin{eqnarray}
\label{eqFOU}
f^A(\vartheta,\eta) = \int \frac{d\kappa_- \xx p_-}{2\pi}~\e^{i\kappa_-%
\xx p_- \vartheta}
\langle p_1,-p_2|O^A(\kappa_+,\kappa_-)|p_1,-p_2\rangle 
(\xx p_-)^{1-d_A}~,
\end{eqnarray}
where $d_q=1$ and $d_G=2$. Eq.~(\ref{eqFOU}) and the inverse Fourier 
transform
\begin{eqnarray}
\label{eqFOUI}
\langle p_1,-p_2|O^A(\kappa_+,\kappa_-)|p_1,-p_2\rangle (\xx p_-)^{1-d_A}
= \int_{1/\eta}^{-1/\eta}
d\vartheta'~\e^{-i\kappa_-' \xx p_- \vartheta'} f^A(\vartheta',\eta)
\end{eqnarray}
are used to obtain evolution equations for $f^A(\vartheta,\eta;\mu^2)$.
To perform the $\kappa_-$--integral the dependence on this variable has 
to be made explicit in the anomalous dimensions $\gamma^{AB}$ to all
orders. The scalar operator matrix elements, Eq.~(\ref{eqSCAM}), do not 
depend  on the light cone mark $\kappa_+$ due to translation invariance 
on the  light--cone, which therefore can be set to zero in 
Eq.~(\ref{eqAN1}). Furthermore, the anomalous dimension $\gamma^{AB}$
obeys the re-scaling relation, cf.~\cite{BGR},
\begin{eqnarray}
\label{eqresc}
\gamma^{AB}(\kappa_+,\kappa_-,\kappa_+',\kappa_-';\mu^2) = \sigma^{d_{AB}}
\gamma^{AB}(\sigma\kappa_+, \sigma\kappa_-,\sigma\kappa_+',
\sigma\kappa_-')~,
\end{eqnarray}
with $d_{AB} = 2 + d_A - d_B$. The anomalous dimension transforms to
\begin{eqnarray}
\label{eqAN1}
\int D\kappa'~\kappa_-^{d_B-d_A}
\gamma^{AB}\left(0,1,\frac{\kappa_+'}{\kappa_-},
\frac{\kappa_-'}{\kappa_-};\mu^2\right) &=& 
\int D\alpha~\kappa_-^{d_B-d_A}
\widehat{K}^{AB}(\alpha_1,\alpha_2;\mu^2) \nonumber\\  &\Df&
\kappa_-^{d_B-d_A}
\int_0^1 du (1-u) \int_0^1 d\xi  \nonumber\\
& & ~~~~~~~~~~\times
\widehat{K}^{AB}(\xi(1-u),(1-\xi)(1-u);
\mu^2) \nonumber\\
&=& \kappa_-^{d_B-d_A} \int_0^1 du  \widehat{\widehat{K}}^{AB}(u;
\mu^2)~,
\end{eqnarray}
with  $\alpha_1=[1-(\kappa_+'+\kappa_-')/\kappa_-]/2=\xi(1-u),
\alpha_2=[1-(\kappa_+'-\kappa_-')/\kappa_-]/2=(1-\xi)(1-u)$.
The $\kappa_-$--integral yields
\begin{eqnarray}
\label{eqKAPM}
\int \frac{d\kappa_- \xx p_-}{2\pi}~\e^{i\kappa_- \xx p_-(\vartheta-u
\vartheta')} (\xx p_- \kappa_-)^{d_B - d_A} 
= \widetilde{O}^{AB}(u\vartheta' -
\vartheta) =
\left\{\begin{array}{ll} \delta(u \vartheta' -  \vartheta) &~
{\rm for}~A=B=q,G \\
\partial_u \delta(u \vartheta'-\vartheta) &~
{\rm for}~A=q,~B=G \\
\theta(u \vartheta'-\vartheta)/\vartheta  &~
{\rm for}~A=G,~B=q
\end{array} \right. \nonumber
\end{eqnarray}
The following evolution equations for the distribution functions 
$f^A(\vartheta,\eta)$ are obtained~:
\begin{eqnarray}
\label{eqEV1}
\mu^2 \frac{d}{d\mu^2} f^A(\vartheta,\eta;\mu^2)
= \int_0^1 du \int_\vartheta^{-{\rm sign}(\vartheta)/\eta}
d\vartheta'~\widetilde{O}^{AB}(u\vartheta'-\vartheta)~\widehat{\widehat
{K}}^{AB}(u;\mu^2) f_B(\vartheta',\eta;\mu^2)~.
\end{eqnarray}
The functions
\begin{eqnarray}
\label{eqANOMF}
\vartheta' \int_0^1 du~\widetilde{O}^{AB}(u\vartheta'-\vartheta)
\widehat{\widehat{K}}^{AB}(u;\mu^2) \equiv P^{AB}\left(\frac{\vartheta}
{\vartheta'},\mu^2\right)
\end{eqnarray}
are the {\it forward} splitting functions, which are independent of 
$\eta$ resp. $x_{\PP}$, leading to
\begin{eqnarray}
\label{eqEV2}
\mu^2 \frac{d}{d\mu^2} f^A(\vartheta,\eta;\mu^2)
=  \int_\vartheta^{-{\rm sign}(\vartheta)/\eta}  \frac{d\vartheta'}
{\vartheta'} P^{AB}\left(\frac{\vartheta}{\vartheta'},\mu^2\right)
f_B(\vartheta',\eta;\mu^2)~.
\end{eqnarray}
They act on the momentum fraction $\vartheta$ of the diffractive parton
densities. Note that the range of $\vartheta'$ is yet different from that
of the corresponding quantity for deep inelastic scattering. After
identifying $\vartheta = 2\beta$ taking the absorptive part, however, one
arrives at the twist--2 evolution equation
\begin{eqnarray}
\label{eqEV4}
\mu^2 \frac{d}{d\mu^2} f^{D}_A(\beta,x_{\PP};\mu^2)
= \int_\beta^1 \frac{d\beta'}{\beta'} P_{A}^B\left(\frac{\beta}{\beta'};
\mu^2\right)
f_B^D(\beta',x_{\PP};\mu^2)~.
\end{eqnarray}
In the case of deep inelastic
scattering the corresponding identification for the momentum fraction is
$z = x$. The dependence of the diffractive parton densities w.r.t. $\eta$,
or $x_{\PP}$, is {\it entirely} parametric and not changed under the 
evolution, which affects $\beta$. In the case of the twist--2 
contributions factorization proofs were 
given~\cite{FACT}\footnote{They apply correspondingly to the kinematic
of fracture functions, see \cite{TV} and for an analysis in $\Phi^3$
theory in $d=6$ dimensions \cite{GRAZ}.}, leading
to the same evolution equations,~(\ref{eqEV4}), which are confirmed by
the present derivation. For phenomenological applications to 
next-to-leading order, see e.g.~\cite{RSBJP}.

We expressed the Compton amplitude with the help of the light--cone 
expansion at short distances and applied this representation to the
process of deep--inelastic diffractive scattering using Mueller's
generalized optical theorem. This representation is {\it not} limited to
leading twist operators but can be extended to all higher twist
operators synonymously. The corresponding evolution equations for
the higher twist hadronic matrix elements, which depend on more momentum
fractions $\vartheta_i$ than one, transform analogously to the case
of twist--2 being outlined above. A central parameter 
$\kappa_+ = \frac{1}{n} \sum_{i=1}^n \kappa_i$ may be set to zero, and 
analogous re--scaling relations apply. By virtue of this also here 
forward evolution equations are derived. However, the connection of the 
momentum fractions $\vartheta_i$ to the outer kinematic parameters is 
less trivial than in the case of twist--2 due to the structure of the 
corresponding Wilson coefficients. This is the case also for deep 
inelastic scattering, cf. e.g.~\cite{BRRZ}.
\section{Conclusions}
\label{sec-5}

\vspace{1mm}
\noindent
The differential cross section of unpolarized deep--inelastic 
$ep$--diffractive
scattering is described by four structure functions for pure photon
exchange, which depend on the four kinematic variables, $x, Q^2, x_{\PP}$
and $t$. In the limit of vanishing target masses and $t \rightarrow 0$
only two structure functions contribute. In the case of hard diffractive
scattering the scaling violations of these structure functions can be
described perturbatively. This is possible in transforming the 
non--forward amplitude of the process via Mueller's optical theorem, by
rotating the final--state proton into an initial--state anti--proton.
The Compton--amplitude is calculated for the hadronic two--particle state
$\langle p_1,-p_2|$ and w.r.t. this state in the forward direction.
The diffractive parton densities are associated to the general 
two--variable distribution functions, which describe the hadronic matrix
element. The scaling variable of the diffractive parton densities, which
directly compares to the Bjorken variable $x$ in the deep--inelastic case,
is $\beta = x/x_{\PP}$. However, the Callan--Gross relation takes the
usual form $F_2^D(x,Q^2,x_{\PP}) = 2x F_1^D(x,Q^2,x_{\PP})$. Due to
the transformation of the problem, which is made possible applying 
Mueller's optical theorem, the anomalous dimensions are the same as for
forward scattering. We demonstrated this by an explicit calculation
in the case of the twist--2 operators, however, the same mechanism
applies also for higher twist operators using the light cone expansion.
In the case of the twist--2 contributions the effective momentum fraction,
$\vartheta = z_- +z_+/\eta$ may be identified with the variable $2\beta$
for diffractive scattering. The dependence of the parton densities on
$x_{\PP}$ is not affected by QCD--evolution, which acts on the variable
$\beta$.

\vspace{2mm}
\noindent
{\bf Acknowledgment.}~For discussions we would like to thank
J. Bartels, W. Buchm\"uller, J.~Dainton, B.~Geyer, P. Kroll, and G. Wolf.


\begin{thebibliography}{999}
%
\bibitem{EXP1}
M. Derrick et al., ZEUS collaboration, Phys. Lett. {\bf B315} (1993) 481;
\\
T. Ahmed et. al., H1 collaboration, Nucl. Phys. {\bf B429} (1994) 477.
%
\bibitem{EXP2}
J. Breitweg et al., ZEUS collabortaion, Eur. Phys. J. {\bf C6} (1999) 43;
\\
C. Adloff et al., H1 collaboration, Z. Phys. {\bf C76} (1997) 613.
%
\bibitem{FL}
C. Adloff et al., H1 collaboration, Phys. Lett. {\bf B393} (1997) 452.
%
\bibitem{EXP3}
H. Abramowicz and J. Dainton, J. Phys. {\bf G22} (1996) 911.
%
\bibitem{BAKO}
J. Bartels and M. Kowalski, Eur. Phys. J. {\bf C19} (2001) 693.
%
\bibitem{DL}
A. Donnachie and P.V. Landshoff, Phys. Lett. {\bf B191} (1987) 309;
Nucl. Phys. {\bf B286} (1987) 704;  Phys. Lett. {\bf B437} (1998) 408;
{\tt hep-ph/0105088} and references  quoted therein.
%
\bibitem{SUR}
M. W\"usthoff and A.D. Martin, J. Phys. {\bf G24} (1999) R309;\\
A. Hebecker, Phys. Rep. {\bf 331} (2000) 1 and references therein.
%
\bibitem{NP}
H. Navelet and R. Peschanski, {\tt hep-ph/0105030} and references 
therein.
%
\bibitem{INGEL}
G. Ingelman and P. Schlein, Phys. Lett. {\bf B152} (1985) 256;\\
J. Bartels and G. Ingelman, Phys. Lett. {\bf B235} (1990) 175;\\
G. Ingelman, {\tt hep-ph/9912534}.
%
\bibitem{BH}
W. Buchm\"uller and A. Hebecker, Nucl. Phys. {\bf B476} (1996) 293;\\
W. Buchm\"uller, T. Gehrmann, and A. Hebecker, Nucl. Phys. {\bf B537} 
(1999) 477.
%
\bibitem{EBKW}
J. Bartels, J. Ellis, M. Kowalski, and M. W\"usthoff,  
Eur. Phys. J. {\bf C7} (1999) 443.
%
\bibitem{FS}
D.~M\"uller, D. Robaschik, B. Geyer, F.~Dittes, and J.~Ho\v{r}ej\v{s}i,
Fortschr. Phys.  {\bf 42} (1994) 2; \\
X. Ji,  J. Phys. {\bf G24} (1998) 1181.
%
\bibitem{RAD}
A.V.~Radyushkin, Phys. Rev. {\bf D56} (1997) 5524; {\tt hep-ph/0101225}.
%
\bibitem{BGR}
J.~Bl\"umlein, B.~Geyer, and D.~Robaschik, Nucl. Phys. {\bf B560} (1999) 
283.
%
\bibitem{KROLL}                           
M. Diehl, T. Feldmann, R. Jakob, and P. Kroll, Phys. Lett. {\bf B460}
(1999) 204; Eur. Phys. J. {\bf C8} (1999) 409;\\
P. Kroll, Nucl. Phys. {\bf A666} (2000) 3.
%
\bibitem{LAT}
G. Martinelli and C.T. Sachrajda, Phys. Lett. {\bf B190} (1987) 151;
{\bf 196} (1987) 184;\\
M. G\"ockeler et al., Nucl. Phys. (Proc. Suppl.) {\bf 53} (1997) 81;\\
M. Guagnelli, K. Jansen, and R. Petronzio, Phys. Lett. {\bf B459} (1999)
594; {\bf B493} (2000) 77.
%
\bibitem{MD}
M. Diehl, Nucl. Phys. {\bf B} (Proc. Suppl. {\bf 79} (1999) 723, Proc.
DIS 99, eds. J. Bl\"umlein  and T. Riemann.
%
\bibitem{TARR}
B. Tarrach, Nuov. Cim. {\bf 28A} (1975) 409.
%
\bibitem{COL}
J.C. Collins, {\tt hep-ph/9705393}.
%
\bibitem{BR}
J. Bl\"umlein and D. Robaschik, Nucl. Phys. {\bf B581} (2000) 449.
%
\bibitem{AHM}
A.H. Mueller, Phys. Rev. {\bf D2} (1970) 2963; Phys. Rev. {\bf D4} (1971)
150;\\
P.D.P. Collins, {\sf An Introduction to Regge Theory and High Energy
Physics}, (Cambridge University Press, Cambridge, 1977), pp.~331.
%
\bibitem{PARA}
S. Weinberg, {\sf The Quantum Theory of Fields}, {\bf Vol.~1}, (Cambridge
University Press, Cambridge, 1995), pp.~452;\\
J. Bl\"umlein, J. Eilers, B. Geyer, and D. Robaschik, to appear.
%
\bibitem{NFLO}
T.~Braunschweig, B.~Geyer, and D.~Robaschik, Ann. Phys. (Leipzig) 
{\bf 44} (1987) 407;\\
I.I.~Balitsky and V.M.~Braun, Nucl. Phys. {\bf B311} (1988/89) 541;\\
A.V.~Radyushkin,
Phys. Lett. {\bf B385} (1996) 333; Phys. Rev. {\bf D56} (1997) 5524;\\
J.~Bl\"umlein, B.~Geyer, and D.~Robaschik, Phys. Lett. {\bf B406} (1997) 
161 and Erratum;\\
I.I. Balitsky and A.V. Radyushkin, Phys. Lett. {\bf B413} (1997) 114;\\
J.~Bl\"umlein, B.~Geyer, and D.~Robaschik, {\tt hep-ph/9711405}, in~:
Proc. of the Workshop~{\sf Deep Inelastic Scattering Off Polarized
Targets: Theory Meets Experiment}, eds.~J.~Bl\"umlein et al., (DESY,
Hamburg, 1997) DESY 97--200, p.~196.\\
X.~Ji, Phys. Rev. Lett. {\bf 78} (1997) 610; Phys. Rev. {\bf D55} (1997)
7114;\\
L. Mankiewicz, G. Piller, and T. Weigl, Eur. J. Phys. {\bf C5} (1998) 
119.
%
\bibitem{NFNLO}
F.M. Dittes and A.V. Radyushkin, Phys. Lett. {\bf B134} (1984) 359;\\
S.V. Mikhailov and A.V. Radyushkin, Nucl. Phys. {\bf B254} (1985) 89;\\
F.M.~Dittes, D.~M\"uller, D. Robaschik, B. Geyer, and J.~Ho\v{r}ej\v{s}i,
Phys. Lett. {\bf B209} (1988) 325;\\
A.V. Belitsky, D. M\"uller, L. Niedermaier, and A. Sch\"afer,
Nucl. Phys. {\bf B546} (1999) 279.
%
\bibitem{FACT}
A. Berera and D.E. Soper, Phys. Rev. {\bf D50} (1994) 4328; {\bf D53}
(1996) 6162;\\
J. Collins, Phys. Rev. {\bf D57} (1998) 3051; Erratum: {\bf D61} (2000)
019902;\\
F. Hautmann, Z. Kunszt, and D.E. Soper, Nucl. Phys. {\bf B563} (1999) 
153.
%
\bibitem{TV}
L. Trentadue  and G. Veneziano, Phys. Lett. {\bf B323} (1994) 201.
%
\bibitem{GRAZ}
M. Grazzini, Phys. Rev. {\bf D57} (1998) 4352.
%
\bibitem{RSBJP}
C. Royon, L. Schoeffel, J. Bartels, H. Jung, and R. Peschanski, 
Phys. Rev. {\bf D63} (2001) 074004.
%
\bibitem{BRRZ}
R.K. Ellis, W. Furmanski, and R. Petronzio, Nucl. Phys. {\bf B212}
(1983) 29;\\
J. Bl\"umlein, V. Ravindran, J. Ruan, and W. Zhu, Phys. Lett. {\bf B594}
(2001) 235.
\end{thebibliography}
\end{document}